\documentclass[12pt]{article}

\usepackage{epsfig}
\usepackage{amssymb}
\usepackage{graphics}
\usepackage{bbm}
\let\a=\alpha   \let\b=\beta

\newcommand{\beq}{\begin{equation}}
\newcommand{\eeq}{\end{equation}}
\newcommand{\beqn}{\begin{eqnarray}}
\newcommand{\eeqn}{\end{eqnarray}}

\newcommand{\nn}{\nonumber}

\newcommand{\be}{\begin{equation}}
\newcommand{\ee}{\end{equation}}
\newcommand{\ba}{\begin{eqnarray}}
\newcommand{\ea}{\end{eqnarray}}
\newcommand{\bdm}{\begin{displaymath}}
\newcommand{\edm}{\end{displaymath}}

\def\b{\beta}

\def\a{\alpha}

\newcommand{\ie}{{\it i.e.\ }}
\newcommand{\eg}{{\it e.g.\ }}

\DeclareMathAlphabet{\mathpzc}{OT1}{pzc}{m}{it}
\def\bea{\begin{eqnarray}}
\def\eea{\end{eqnarray}}
\def\beas{\begin{eqnarray*}}
\def\eeas{\end{eqnarray*}}
\def\sla{\raise.15ex\hbox{$/$}\kern-.57em}

\def\bea{\begin{eqnarray}}
\def\eea{\end{eqnarray}}

\def\sla{\raise.15ex\hbox{$/$}\kern-.57em}
\def\ie{{\it i.e.}~}
\def\eg{{\it e.g.}~}

\def\a{\alpha}
\def\b{\beta}

\def\cA{{\cal A}}

\def\cH{{\cal H}}
\def\cI{{\cal I}}

\def\cK{{\cal K}}

\def\cM{{\cal M}}
\def\cN{{\cal N}}

\def\cQ{{\cal Q}}

\def\cT{{\cal T}}

\def\cX{{\cal X}}

\newcommand{\ft}[2]{{\textstyle\frac{#1}{#2}}}
\begin{document}
\begin{titlepage}
\rightline{ROM2F/2009/02, \, NSF-KITP-09-13}
\vskip 2cm
\begin{center}
{\Large\bf From Twists and Shifts to L-R asymmetric
D-branes\footnote{Talk delivered at the 4-th RTN {\it
``Forces-Universe''} EU network Workshop in Varna,  September
2009.}}
\end{center}
\vskip 2cm
\begin{center}
{\large\bf Massimo Bianchi}\\~\\
{\sl Dipartimento di Fisica and Sezione I.N.F.N. \\ Universit\`a di Roma ``Tor Vergata''\\
Via della Ricerca Scientifica, 00133 Roma, Italy}\\
and \\ {\sl Kavli Institute for Theoretical Physics \\
University of California, Santa Barbara, CA 93106-4030}
\end{center}
\vskip 3.0cm
\begin{center}
{\large \bf Abstract}
\end{center}

In the first part of the talk, we discuss non-geometric twists and
shifts and briefly review asymmetric orbifolds and free fermion
constructions. These allow us to build Type IIB models with $\cN =
1_L + 1_R$ and $\cN = 1_L + 0_R$ models having few or no moduli.
We then consider unoriented projections of the former and
(`exotic') D-branes in the latter.

In the second part, devoted to L-R asymmetric D-branes, we review
how extended supergravity vacua in $D=4$ can be embedded in Type
II superstrings. We then identify bound states of D-branes with
residual susy and non-trivial R-R couplings. We discuss the
$\cN=6=2_{_L} + 4_{_R}$ case in detail and sketch other extended
susy cases. Finally we describe the resulting open string
excitations.

We conclude with some speculations and possible interesting
developments.



\vfill

\end{titlepage}


\section{Foreword}

So far, the combined effect of twists and shifts has not been
systematically explored in the context of (unoriented) strings
\cite{BPS}. The basic idea is that chiral twists can freeze out
untwisted moduli, while (orthogonal) non-geometric shifts prevent
massless twisted moduli from appearing.

In this way, one may provide exact CFT descriptions of `T-folds'
\cite{Tfolds} in terms of (a)symmetric orbifolds
\cite{Narain:1986qm, Vafa:1986wx} and/or free fermions
\cite{Kawai:1986va, abk} and allow for a systematic search of
perturbative vacua with few moduli. Moreover they suggest the
existence of `new' kinds of L-R asymmetric D-branes
\cite{Bianchi:2008cj} with largely unexplored phenomenological
applications.

In the era of LHC, taming superstring vacua with few or no moduli
is more than necessary \cite{Lust:2008qc, Antoniadis:2007uz}.

The talk is divided into two parts.

In the first part, based on \cite{Anastasopoulos:2009kj}, we will
discuss twists and shifts, briefly review asymmetric orbifolds
\cite{Narain:1986qm, Vafa:1986wx} and free fermions
\cite{Kawai:1986va, abk}, build Type IIB models with $\cN = 1_L +
1_R$ and $\cN = 1_L + 0_R$ from $T^6_{SO(12)}$, find consistent
unoriented projections and (`exotic') D-branes
\cite{Bianchi:1999uq}

In Part II, devoted to L-R asymmetric D-branes and based on
\cite{Bianchi:2008cj}, we will discuss extended supergravity vacua
in string theory \cite{Ferrara:1989nm}, identify bound states of
D-branes with residual susy and non-trivial R-R couplings, focus
on the $\cN=6=2_{_L} + 4_{_R}$ case, sketch other extended susy
cases, and describe the resulting open string excitations.

We will end with an outlook.

\section{(Unoriented) T-folds with few or no T's}

\subsection{Asymmetric orbifolds and free
fermions}

Asymmetric orbifolds \cite{Narain:1986qm, Vafa:1986wx} are
constructions where Left- and Right-moving fields on the string
worldsheet are treated differently or are different altogether.
Stringent modular invariance constraints must be satisfied.

For simplicity, consider $Z_{2L}$ chiral reflections of
Left-moving internal bosonic and fermionic coordinates \bea
 \cI_i: && X_L^i   \rightarrow  - X^i_L \ ,
 \quad \quad X_R^i
\rightarrow  X^i_R \ , \quad \quad
   \psi^i \rightarrow -
\psi^i \ ,  \quad \quad \tilde\psi^i \rightarrow  \tilde \psi^i \
.
 \nn \eea
Similarly one can act on the Right-movings.

In a $Z_{2L}\times Z_{2L}'\times Z_{2R}\times Z_{2R}'$ orbifold
with generators $\cI_{3456}, \cI_{1256}$ and $ \bar
\cI_{3456},\bar \cI_{1256}$ all untwisted moduli fields, except
the axio-dilaton are frozen \cite{Anastasopoulos:2009kj}.

If one combines chiral twists with chiral shifts along orthogonal
directions such as
$$
 \sigma_{j}: X^j_L \to X^j_L+\delta^j \ , \qquad  X^j_R \to X^j_R \quad ;
$$
 with $2\delta$ a lattice vector, most massless twisted
 moduli are prevented from appearing. Clearly such duality twists
 and shifts are symmetries of very special tori, in particular
 those based on free fermions \cite{Kawai:1986va, abk, Ferrara:1989nm}. In these
constructions one `fermionizes' the internal coordinates, so that
the worldsheet supercurrent becomes \cite{ABKW}
$$G=\Psi^\mu \partial X_\mu + \psi^i y^i w^i$$

Preservation of $G$, up to a sign, under parallel transport along
non-contractible cycles requires that the boundary conditions of
the various fermions be related. For `real' fermions, \ie $Z_2$
twists, one has b.c. ($\Psi^\mu$) = b.c. ($\psi^i$) + b.c. ($y^i$)
+ b.c. ($w^i$), $\forall \ i $ and $\mu$.

The construction is specified by the choice of basis sets of
fermions $b_\a$. Modular invariance (or level matching) imposes
stringent constraints. Focussing on $Z_2$ twists one finds
\cite{abk} \bea
   n_{_{L-R}}(b_\alpha)&=&0~{\rm mod }~8 \ ;\nn\\
    n_{_{L-R}}(b_\alpha \cap b_\beta)&=&0~{\rm mod }~4 \ ;\nn\\
    n_{_{L-R}}(b_\alpha \cap b_\beta \cap b_\gamma)&=&0~{\rm mod }~2 \ ;\nn\\
       n_{_{L-R}}(b_\alpha \cap b_\beta\cap b_\gamma \cap b_\delta)&=&0~{\rm mod }~2 \ ;
  \nn \eea
    where $n_{_{L-R}}(b) = n_{_L}(B) - n_{R}(b)$ denotes the difference
    between the numbers of L- and R- moving fermions in set $b$.
Notice that combinations such as $y^iw^i$ in any basis set act as
non-geometric shifts.

 In what follows, our starting point will be
the Type IIB superstring on the $T^6$ maximal torus of $SO(12)$
with $\cN = 8 = 4_L+4_R$. The presence of a non vanishing B-field
plays a subtle crucial role in the reduction of the number of
twisted sectors \cite{CRISTINA}. For our purposes, the fermionic
description of the model relies on the choice of $$ F = \{
\psi^{1\ldots 8} \, y^{1\ldots 6} \, w^{1\ldots 6}  | \,
\tilde\psi^{1\ldots 8}\, \tilde{y}^{1\ldots 6}\,\tilde{w}^{1\ldots
6}   \} \ , \quad \quad S = \{\psi^{1\ldots 8} \}\ , \quad\quad
\tilde{S} = \{\tilde \psi^{1\ldots 8}  \} $$ as basis sets. The
one-loop `torus' partition function reads
$$ \cT_{(4,4)} = |V_8 - S_8|^2 (|O_{12}|^2 +|V_{12}|^2 +|S_{12}|^2
+|C_{12}|^2) $$ where $O,V,S,C$ denote characters characters of
$SO(2n)$ current algebra \cite{MBthesis} that can be expressed in
terms of Jacobi $\vartheta$ functions.

Alternatively, keeping only $F$ and $S$, one finds an $\cN = 4 =
4_L+0_R$ model with $SU(2)^6$ gauge group and $SO(20)$
`pseudo-symmetry' \cite{Dixon:1987yp}. Its partition function
reads
$$ \cT_{(4,0)} = (V_8 - S_8)
(O_{12}\bar{V}_{20} +V_{12}\bar{O}_{20} - S_{12}\bar{S}_{20} -
C_{12}\bar{C}_{20}) $$

There has been recent revival of interest on free fermion
constructions both for heterotic strings, where spinor/vector
duality has been `established' and realistic models proposed, as
well as for Type II strings where (non) magic hyper-free
supergravities have been constructed and models with few (twisted)
moduli found \cite{CRISTINA, Donagi:2008xy}.

 Our analysis, aimed at finding unoriented T-folds with few or no
 T's, was largely motivated by two seemingly unrelated
 investigations the { CDMP model} \cite{Camara:2007dy} and {DJK  `minimal'
 model} \cite{Dolivet:2007sz}.

The {CDMP model}, based on a previous observation in
\cite{Vafa:1994rv}, is a standard geometric freely acting orbifold
$T^6/Z_2 \times Z_2$, that yields Type I / Heterotic dual pairs.
All twisted moduli are massive. Only untwisted moduli $T_I, U_I$
survive. Including gaugino condensate(s) in the open string sector
and/or 3-form fluxes \cite{flux} allows to partially stabilize the
dilaton and other moduli.

The {DJK `minimal' model} \cite{Dolivet:2007sz}, a (non magic
\cite{MAGIC}) hyper-free Type II model with $\cN = 2 = 2_R + 0_L$
susy, is a fermionic construction based on the choice of sets: $F,
S, \bar{S}$, $\bar{b}_1$, $ b_1 = \{\Psi^\mu, \psi^{1,2};
y^{3,4,5,6}, y^1w^1 | \bar{y}^5\bar{w}^5 \}$, ${b}_2 =
\{\Psi^\mu,\psi^{3,4}; {y}^{1,2}, {w}^{5,6}
 {y}^3{w}^3 | \bar{y}^6\bar{w}^6 \}$,
 ${b}_3 = \{\Psi^\mu, \psi^{5,6};
{w}^{1,2,3,4} y^6 w^6|
 \bar{y}^6\bar{w}^6 \}$

Only the dilaton vector (!) multiplet survives. All susy
associated to Left-moving supercharges are broken, $\cN_{_L}=0$,
due to the L-R asymmetric $(-)^{F_{_L}}\sigma$ freely acting
orbifold projection of $T^6_{SO(12)}$.

\subsection{The `minimal' model with ``$h_{11}$'' =``$h_{11}$''=1}

In the perspective of unoriented projections, one is lead to
replace $\bar{b}_3$ with ${b}_2$ and get a L-R symmetric
asymmetric orbifold. Geometric (freely acting) projections
associated to $b_1 \bar{b}_1$ and $b_2 \bar{b}_2$ are combined
with non geometric (freely acting) projections associated to $b_1,
b_2, ... b_1\bar{b}_2$. The surviving spacetime susy is
$\cN_L=\cN_R=1, \cN_{tot} = 2 $. All untwisted moduli except the
dilaton hypermultiplet are projected out. One could hope that all
twisted sectors be massive, thanks to the shifts. Alas, massless
multiplets arise from the $b_1 b_2\bar{b}_1 \bar{b}_2$ twisted
sector, that contributes one hyper and one vector multiplet. By
analogy with `geometric' CY-like compactifications one is lead to
introduce effective Hodge numbers and get
``$h_{11}$''=``$h_{21}$''=1 in this case.

As we will see later, unoriented projections can produce
$1_u+2_t-n$ chiral-plets and $n$ vector-plets ($n=0,1$).
Unfortunately, at a first scan, the open string spectrum seems to
be non chiral and the Chan-Paton group suffers rank reduction due
to the non-vanishing but quantized B-field \cite{BPStor, MBtor,
EWtor, Angelantonj:1999xf, CBetal, Pesando:2008xt}. A systematic
study is under way. MSSM embeddings in similar fermionic
constructions have been recently discussed in
\cite{Kiritsis:2008mu}.

Let us now, give some details of the ${{T}^6}/(Z_{2L}\times
Z_{2L}'\times Z_{2R}\times Z_{2R}')$ model with ``$h_{11}$'' =
``$h_{21}$'' =1 \cite{Anastasopoulos:2009kj}.

The generators of the orbifold group can be specified by the
following choice of fermionic sets
  \bea
  b_{1}&=& \{ \psi_{3456}~; \quad y_{13456}~; \quad ~~ w_{1}\:|~~~~~~~;\quad ~~~~~\tilde y_{5}~;\quad ~~ \tilde w_{5} ~\}\nn\\
  b_{2}&=& \{ \psi_{1256}~; ~~\quad y_{123}~; \quad w_{356}\:|~~~~~~~;\quad ~~~~~\tilde  y_{6}~;\quad ~~ \tilde  w_{6} ~\}\nn\\
 \tilde  b_{1}&=& \{ ~~~~~~~~;\quad  ~~~~y_{5}~;\quad  ~ w_{15} \:| \tilde \psi_{3456}~; \quad \tilde y_{13456}~; \quad  ~~\tilde w_{1} ~  \}\nn\\
  \tilde  b_{2}&=& \{ ~~~~~~~~;\quad  ~~~~y_{6}~;\quad  ~~ w_{6} \:| \tilde \psi_{1256}~; \quad~~ \tilde y_{123}~; \quad \tilde w_{356}    ~\}\
   \nn \eea
 Defining also $b_3 = b_{1}b_{2}$ and $\tilde b_3 = \tilde b_{1}\tilde b_{2}$
   \bea
  b_{3} &=& \{ \psi_{1234}~; \quad ~ y_{2456}~; \quad w_{1356}
  \:|~~~~~~~;\quad ~~~\tilde y_{56}~;\quad ~~\tilde  w_{56} ~ \}\nn\\
   \tilde  b_{3}&=& \{ ~~~~~~~~;\quad  ~~~y_{56}~;\quad  ~ w_{56} \:|
 \tilde \psi_{1234}~; \quad ~ \tilde y_{2456}~; \quad \tilde w_{1356} ~ \}
 \nn \eea
Orbifold group elements can then be expressed as $b_{\a}^m \tilde
b_{\b}^n$ with $\a,\b=1,2,3$, $m,n=0,1$.

The torus partition function $${\cal T}=\ft{1}{16} \{
\sum_{c,d=0}^3\, \rho_{0c}\, \bar \rho_{0d}\, {\bf
\Lambda}_{00,cd}+ \sum_{a,b=0}^3\, (\rho_{a0}\, \bar \rho_{b0}\,
{\bf \Lambda}_{ab,00}+\rho_{aa}\, \bar \rho_{bb}\, {\bf
\Lambda}_{ab,ab}) +   \sum_{a=1}^3 \sum_{b\neq a}^3\, {\bf
\epsilon_{ab}}\, \rho_{ab}\, \bar \rho_{ab}\, {\bf
\Lambda}_{aa,bb} \}$$ where the chiral amplitudes $\rho_{ab}$ \bea
\rho_{00}&=& {1\over \eta^8 } \, (Q_o+Q_v) \quad\quad
\rho_{0h}={1\over  \eta^8 } \, (Q^{(h)}_o-Q^{(h)}_v)\nn\\
\rho_{h0}&=& {1\over  \eta^8 } \, (Q^{(h)}_s+Q^{(h)}_c) \quad\quad
\rho_{hh}=-{i\over  \eta^8 } \, (Q^{(h)}_s-Q^{(h)}_c)\nn \\
\rho_{hh'}&=& -{i\over \eta^8 } \, (Q^{(h)}_{s-'}-Q^{(h)}_{c-'})
\quad h\neq h' \quad h=1,2,3 \nn\eea can be expressed in terms of
the super-characters
$$Q_o = V_4O_4-S_4S_4 \quad , \quad Q_v = O_4V_4-C_4C_4$$
$$Q_s = O_4S_4-C_4O_4\quad , \quad Q_c
= V_4C_4-S_4V_4$$ while the lattice sums for $h=1,2$ as well as
for $h'=3$ can be conveniently written in terms of $\vartheta$
functions \bea {\bf \Lambda}_{30,30} &=&   {\bf \Lambda}_{03,03}^*
  =\,\ft{i}{2} \left(  \vartheta_4^2\vartheta_2^4 \bar \vartheta_4^4 \bar \vartheta^2_4
   -\vartheta_4^4\vartheta_2^2 \bar \vartheta_4^2 \bar \vartheta_2^4
   \right)\quad
  {\bf \Lambda}_{00,h0} =   {\bf \Lambda}_{00,0h}^*
  =\,\ft12 \left(  \vartheta_3^3\vartheta_4^3 \bar \vartheta_3^5 \bar \vartheta_4
   + \vartheta_3^3\vartheta_4^3 \bar \vartheta_3 \bar \vartheta_4^5 \right)\nn\\
     {\bf \Lambda}_{h0,00} &=&   {\bf \Lambda}_{0h,00}^*
  =\,\ft12 \left(  \vartheta_3^3\vartheta_2^3 \bar \vartheta_3^5 \bar \vartheta_2
   + \vartheta_3^3\vartheta_2^3 \bar \vartheta_3 \bar \vartheta_2^5
   \right)\quad
  {\bf \Lambda}_{h0,h0} =   {\bf \Lambda}_{0h,0h}^*
  =\,\ft{i}{2} \left( \vartheta_4^3\vartheta_2^3  \bar \vartheta_2^5 \bar \vartheta_4
   - \vartheta_2^3\vartheta_4^3 \bar \vartheta_2\bar \vartheta_4^5 \right)\nn\\
  {\bf \Lambda}_{00,30} &=&   {\bf \Lambda}_{00,03}^*
  =\,\ft12 \left(  \vartheta_3^2\vartheta_4^4 \bar \vartheta_3^4 \bar \vartheta^2_4
   + \vartheta_3^4\vartheta_4^2 \bar \vartheta_3^2 \bar \vartheta_4^4
   \right)\quad
     {\bf \Lambda}_{30,00} =   {\bf \Lambda}_{03,00}^*
  =\,\ft12 \left(  \vartheta_3^2\vartheta_2^4 \bar \vartheta_3^4 \bar \vartheta^2_4
   + \vartheta_3^4\vartheta_2^2 \bar \vartheta_3^2 \bar \vartheta_2^4 \right)\nn\\
   {\bf \Lambda}_{00,00} &=& \ft12 (|\vartheta_2|^{12}+ |\vartheta_3|^{12}+ |\vartheta_4 |^{12})
  \quad \quad \quad
      {\bf \Lambda}_{other} =   0
 \nn \eea

As anticipated, the only massless twisted states come from the
$b_3\tilde b_3$ sector. The whole massless spectrum is coded in
 $$
 {\cal T} = |V-S-C|^2+|2O-S-C|^2+\ldots $$
and consists in the
 ${\cal N}=2$ supergravity multiplet coupled to 2 hypers and 1
 vector, \ie ``$h_{11}$''= ``$h_{21}$'' = 1.

\subsection{Models with
$\cN=1_L + 1_R$ based on $T^6_{SO(12)}$}

 We then performed a systematic search of models with
 $\cN=1_L + 1_R$ from $T^6_{SO(12)}/Z_2^4$, corresponding to the
 choice of basis sets $F,S,\tilde{S}$ plus four more sets of the form
 \bea &&
b_1 = I_{3456}\, \sigma^{i_1 i_2 \ldots }\,\bar \sigma^{k_1 k_2 \ldots } = \{ (\psi \,y)^{3456}  \, (y\, w)^{i_1 i_2 \ldots }  | (\tilde y\, \tilde w)^{k_1 k_2 \ldots }  \} \ , \nn\\
&&b_2 = I_{1256}\, \sigma^{j_1 j_2 \ldots }\,\bar \sigma^{l_1 l_2 \ldots } = \{(\psi\, y)^{1256}  \, (y\, w)^{j_1 j_2 \ldots }  | (\tilde y\, \tilde w)^{ l_1 l_2 \ldots }   \} \ , \nn\\
&&\bar{b}_1= \bar I_{3456}\, \sigma^{k_1 k_2 \ldots }\,\bar
\sigma^{i_1  i_2 \ldots }  = \{ ( y\,  w)^{k_1 k_2 \ldots }  |
(\tilde\psi\, \tilde y)^{3456}
  (\tilde y\, \tilde w)^{i_1 i_2 \ldots }     \} \ , \nn\\
&&\bar{b}_2 = \bar I_{1256}\, \sigma^{l_1 l_2 \ldots }\,\bar
\sigma^{ j_1  j_2 \ldots } = \{ ( y\,  w)^{ l_1 l_2 \ldots } |
(\tilde\psi\, \tilde y)^{1256} (\tilde y\, \tilde w)^{j_1 j_2
\ldots }  \} \ , \nn \eea As above, in view of the resulting $\cN
= 1_L + 1_R$ supersymmetry, one can introduce ``effective'' Hodge
numbers ``$h_{11}$'' = $n_h$ -1, ``$h_{21}$'' = $n_v$. We found
three finite sequences of models with (``$h_{11}$'',``$h_{21}$''):
\bea
&&(n,n),\: n=1,2,3,4,5,9,\quad {\rm self-mirror},\quad \chi =0 \nn\\
&&(2n,2n+6)/(2n+6,2n),\: n=0,1,2,\quad {\rm mirror \, pairs}, \quad \chi =\mp 12 \nn\\
&&(2n+3,2n+15)/(2n+15,2n+3),\: n=0,1,\quad {\rm mirror \, pairs},
\quad \chi =\mp 24 \nn \eea

Once again, the non vanishing but quantized B-field plays a subtle
role in reduction of twisted sectors \cite{CRISTINA} while
discrete torsion \cite{Vafa:1986wx, Vafa:1994rv}, \ie relative
signs between disconnected orbits of the modular group in the
one-loop partition function, lead to the exchange of vectors and
hypers, \ie generalized mirror symmetry for these non-geometric
Type II vacuum configurations. It is amusing to see that $\chi =
12k$ in all cases we analyzed. We have no convincing explanation
for this except for the obvious fact that the models at hand can
be thought of as non-geometric orbifolds of $K3\times T^2$.

\subsection{Models with
$\cN=1_L + 0_R$ based on $T^6_{SO(12)}$}

In the same vein, one can systematically scan for models with
$\cN=1_L + 0_R$ based on $T^6_{SO(12)}$. To this end, one keeps
only the sets $F$ and $S$ (not $\tilde{S}$) plus, for instance,
two more sets $b_1$ and $b_2$. The latter spacetime susy to
$\cN=1_L+ 0_R $ and the internal (pseudo)symmetry $SO(20)$. The
``true'' gauge symmetry can only be determined after a careful
analysis of the vertex operators for vectors and their OPE's. Not
unexpectedly, it turns out to be a subgroup of $SU(2)^6$ with
abelian factors and the possibility of further (perturbative)
Higgs mechanism.

The massless spectrum can be decomposed according to  $\cT_0={\bf
G}_{1} +n_v{\bf V}_{1} +n_{v'}{\bf V}'_{1} +n_c {\bf C}_{1} +
n_{c'} {\bf C}'_{1} $ \bea {\rm with} \quad
{\bf G}_{1}   + {\bf C}_{1}  &=& (V-S-C) \,\bar V  \ , \nn\\
 {\bf V}_{1} &=& (V-S-C) \,\bar O \qquad
 {\bf V}'_{1} =  (S-O) \, \bar S + (C-O) \,\bar C \ , \nn\\
{\bf C}_{1} &=& (2O-S-C) \,\bar O
  \qquad
{\bf C}'_{1} =  (S-O) \, \bar C + (C-O) \,\bar S \ , \nn
  \eea As indicated there are two different kinds of chiral and vector
multiplets distinguished by the nature, NS-NS or R-R, of the
bosonic components. Though L-R asymmetric, these models allow the
introduction of generalized D-branes that couple to (twisted) R-R
states \cite{Bianchi:2008cj}.

The simplest example we found of a model with $\cN=1_L + 0_R$
spacetime susy has  $(n_v,n_v';n_c,n_c ')=(14, 0;5, 0)$ and
correspond to the choice of orbifold generators
$$b_{1}=  I _{3456} ~ \sigma_{12} ~   \overline{\sigma}_{45}  \quad ,
\quad b_{2}=  I _{1256} ~ \sigma_{36} ~  \overline{\sigma}_{5} \ .
$$ In the absence of `exotic' D-branes, the resulting gauge
symmetry, associated to worldsheet currents, is $SU(2)^4\times
U(1)^2$. The (pseudo)symmetry is broken according to $$
O(12)_L\times O(20)_R\to \left[ O(4)^2\times O(2)^2\right]_L\times
\left[ O(2)^2\times O(16)\right] _R  $$ or, rather, $O(16)_R
\rightarrow O(2)\times O(14)$, with $O(2)$ little group in $D=4$.

We also found other $\cN=1_L + 0_R$ Type II models with
$(n_v,n_v';n_c,n_c ')= (10,0;25,0), (8,0;27,0), (6,8;13,8),
(6,8;29,8)$. As apparent, the scan was neither very systematic nor
very inspiring. Yet one should explore the possibility of adding
`exotic' D-branes.

\subsection{Unoriented projections and open strings}

L-R symmetric though non geometric models with $\cN=1_L + 1_R$
admit $\Omega$ projections $$    {\rm Tr}_{\cH_L\otimes \cH_R } \,
    \Omega \,(g^L \otimes g^R)=  {\rm Tr}_{\cH_L} \, g_{_\Omega}  \ ,
     $$
$g_{_\Omega}$  diagonal action  $g^L  g^R$ with Left- and Right-
moving fields identified, i.e. $\bar I_i \to I_i$, $\bar \sigma_i
\to \sigma_i$. In general $g_{_\Omega}$ amplitudes are not chiral
`square roots' of amplitudes in torus partition function. Several
closed string moduli are odd under $\Omega$
 and are thus projected out. D-branes may be needed to cancel R-R
 tadpoles \cite{BPS, MBthesis, Dai:1989ua, Polchinski:1995mt, Gimon:1996rq}.

An unoriented (`Type I') model without open strings can be
constructed from the Type II with $\cN=1_L+1_R$ model with
(``$h_{11}$''= ``$h_{21}$'')=(1,1). Two allowed Klein-bottle
projections are of the form
$$\cK = \frac{1}{16} \sum_{a,b,c,d} {\rm Tr}_{\cH_{Lc}\otimes \cH_{Rd} }
\Omega\, b_{a}\, \bar b_b
 = \frac{1}{4} \sum_{a,b}{\rm Tr}_{\cH_{La}}  b_{b\Omega}= \frac{1}{4 \ \eta^8}  \sum_{a,b=0}^3
\epsilon_{a,b} \, \rho_{ab} \, {\Lambda}[^a_{b}] $$ with
${\Lambda}[^0_{0}] = \vartheta_3^6 +\epsilon\, \vartheta_2^6$ and
\bea &&{\Lambda}[^0_{h}]  = \vartheta_4^3 \vartheta_3^3
+\epsilon\, \vartheta_1^3\vartheta_2^3 , \quad  {\Lambda}[^h_{0}]
= \vartheta_2^3 \vartheta_3^3 +\epsilon\,
\vartheta_3^3\vartheta_2^3 , \quad {\Lambda}[^h_{h}] =
\vartheta_1^3 \vartheta_3^3 + \epsilon\,\vartheta_4^3\vartheta_2^3
\nn \\
&&{\Lambda}[^0_{3}]  = \vartheta_4^2 \vartheta_3^4 +\epsilon\,
\vartheta_1^2\vartheta_2^4, \quad {\Lambda}[^3_{0}]  =
\vartheta_2^2 \vartheta_3^4 +\epsilon\, \vartheta_3^2\vartheta_2^4
, \quad {\Lambda}[^3_{3}]  = \vartheta_1^2 \vartheta_3^4
+\epsilon\, \vartheta_4^2\vartheta_2^4 \nn\eea For $\epsilon=-1$,
$\epsilon_{a,b}=1$, no massless untwisted or twisted tadpoles
appear in the transverse channel and thus no D-branes are needed.
The resulting Type I model consists of closed strings only with
`minimal' $\cN=1$ content $\ft12(\cT+\cK)_{\rm massless}=  {\bf
G}_{1}+  2\, {\bf C}_{1}$

Let us now discuss the simplest unoriented model with open
strings. For simplicity, consider $T^6/Z_{2L}\times Z_{2L}'\times
Z_{2R}\times Z_{2R}'$ with no shifts at the $SO(12)$ point. The
spectrum can be coded in a basis of 64 super-characters, that
appear in the geometric case as well \cite{MBthesis,
Berkooz:1996dw}. There are two choices of signs (discrete
torsion), leading to two different Klein bottle projections. Yet,
they both lead to the same non-chiral massless open string
spectrum, encoded in $( \cA + \cM)/2$:
 ${\cal N}=4$ SYM with gauge group $U(N)\times U(4-N)$. Rank reduction is due to quantized B-field
 \cite{BPStor, MBtor,
EWtor, Angelantonj:1999xf, CBetal, Pesando:2008xt}. Alas, there
seems to be some tension between chirality and moduli
stabilization. Before drawing too drastic conclusions one should
wait for a more systematic analysis of D-branes in non geometric
compactifications of the above kind.

\section{Bound states of L-R asymmetric D-branes}

\subsection{Type II superstring vacua with extended susy}

Type II superstrings can give rise to (non) geometric vacua with
extended supergravity  \cite{Ferrara:1989nm, Dabholkar:1998kv}
\bea\cN = 8
\quad &\leftrightarrow&\quad \cN_{_L} = 4 \ , \ \cN_{_R} = 4 \nn\\
\cN = 6 \quad &\leftrightarrow& \quad \cN_{_L} = 2 \ , \ \cN_{_R}
= 4 \nn\\ \cN = 5 \quad &\leftrightarrow& \quad \cN_{_L} = 1 \ , \
\cN_{_R} = 4 \nn\\ \cN = 4 \quad &\leftrightarrow& \quad \cN_{_L}
= 2 \ , \ \cN_{_R} = 2 \quad {\bf or}\quad \cN_{_L} = 0 \ , \
\cN_{_R} = 4 \nn\\ \cN = 3 \quad &\leftrightarrow& \quad \cN_{_L}
= 1 \ , \ \cN_{_R} = 2 \nn\\ \cN = 2 \quad &\leftrightarrow& \quad
\cN_{_L} = 1 \ , \ \cN_{_R} = 1 \quad {\bf or}\quad \cN_{_L} = 0 \
, \ \cN_{_R} = 2 \nn\eea

In the spirit of the first part of the talk, asymmetric orbifolds
and free fermions, \ie twists and shifts, can provide `exact'
(rational) CFT descriptions of the above. For our present purpose
it is crucial to observe that, even if L-R asymmetric, whenever
massless R-R states (\eg graviphotons) survive there must be
bound-states of D-branes they couple to \cite{Dai:1989ua,
Polchinski:1995mt}. As we will see \cite{Bianchi:2008cj}, these
`exotic' D-branes preserve some fraction of extended susy and
satisfy BPS conditions. In many cases, one can resort to CFT
techniques, \ie use boundary states for magnetized D-branes
\cite{Abouelsaood:1986gd, Bianchi:2005sa, Billo:1998vr,
Callan:1987px, Di Vecchia:1997pr, Di Vecchia:2006gg}, to determine
the resulting open string excitations.

\subsection{$\cN = 6 = 2_L + 4_R$ case }

Spontaneous breaking $\cN = 8 \rightarrow \cN =6$ via chiral $Z_2$
twist of the L-movers corresponds to T-duality twists on four
internal directions, $T^4_t$ ) \be X^i_{_L} \rightarrow - X^i_{_L}
\quad , \quad \Psi^i_{_L} \rightarrow -\Psi^i_{_L} \quad , \quad
i=6,7,8,9 \nn\ee accompanied by an order two shift along the
untwisted $T_s^2$ \cite{Ferrara:1989nm}. Unbroken susy's satisfy $
\cQ_{_L} = \Gamma_{6789} \cQ_{_L} $, while no conditions are to be
imposed on $\cQ_{_R}$. After dualizing all masseless 2-forms into
axions, the $30=2_{_{NS-NS}}+12_{_{NS-NS}}+16_{_{R-R}}$ scalar
moduli parameterize the coset space $ \cM^{D=4}_{\cN = 6} =
SO^*(12)/U(6) $. The $16=8_{_{NS-NS}}+8_{_{R-R}}$ vectors together
with their magnetic duals transform according to the ${\bf 32}$
dimensional chiral spinor representation of $SO^*(12)$.

Let us now consider $\cN=6$ BPS conditions for D-brane
bound-states invariant under twist and shift that couple to the
surviving R-R graviphotons and carry the $16_{_{R-R}}=2_{(1|5)} +
4_{(1|3)} + 6_{(3|3)} + 4_{(5|3)}$ R-R charges
 $$q_1^{a} + {1\over 4!}\varepsilon_{ijkl} q_5^{aijkl}
\quad , \quad q_1^i + {1\over 3!}\varepsilon^i{}_{jkl} q_3^{jkl}
\quad , \quad q_3^{aij} + {1\over 2!} \varepsilon^{ij}{}_{kl}
q_3^{akl} \quad , \quad q_5^{abijk} + \varepsilon^{ijk}{}_l
q_3^{abl}
$$
Consider for instance the state consisting of a D5 wrapped along
the twisted $T^4_t\times S_s^1$ and a D1 along the same $S_s^1$.
The susy condition, $\cQ_{_R} = \Gamma_{04}\Gamma_{6789} \cQ_{_L}
= \Gamma_{04} \cQ_{_L}$, leads to a 1/3 BPS
 state. A different analysis applies to BPS states carrying NS-NS charges
\eg the two massive gravitini and their superpartners, carrying
internal generalized KK momentum, form a complex 1/2 BPS
multiplet.

D-branes in `T-folds' have already been studied from different
vantage points \cite{Brunner:1999fj, Bianchi:1999uq,
Gaberdiel:2002jr, Gutperle:2000bf, Kawai:2007qd, Lawrence:2006ma}
but never in a systematic way.

Other $\cN=6$ cases can be studied. First, instead of $Z_2$ one
can perform a $Z_n$ chiral projection on 4 real (2 complex)
super-coordinates as \be (Z^1, Z^2)_{_L} \rightarrow (\omega Z^1,
\omega^{-1} Z^2)_{_L} \quad , \quad (\Psi^1, \Psi^2)_{_L}
\rightarrow (\omega \Psi^1, \omega^{-1} \Psi^2)_{_L} \nn\ee with
$\omega^n =1$. In order to avoid massless twisted states, one has
to combine it with an order $n$ shift along the `untwisted'
directions $(Z^3_{_L};Z_{_R}^i)$. Alternatively, the maximal torus
of $SU(3)^3$ admits a chiral $Z_3$ projection with no shift.
$\cN=5$ supergravity survives in the untwisted sector. The twisted
sector produces an extra massless gravitino multiplet that
completes the spectrum of $\cN=6$ supergravity
\cite{Dabholkar:1998kv}.

\subsection{Other extended susy cases with L$\neq$R}

Let us list the other extended susy cases with some of their
properties
\begin{itemize}

\item{ $\cN =5=1_{_L} + 4_{_R}$, unique massless spectrum,
non-geometric, uncorrected LEEA (like $\cN=6,8$)}

\item{ $ \cN =4=2_{_L} + 2_{_R}$ uncorrected LEEA, variable $N_v$,
(non)geometric, $SL(2)\times SO(6,N_v)$ symmetry}

\item{ $\cN =3=1_{_L} + 2_{_R}$ uncorrected LEEA, variable $N_v$,
non-geometric / fuxes, $U(3,N_v)$ symmetry}

\item{ $\cN =2=1_{_L} + 1_{_R}$, (non) geometric, quantum
corrections absent in special cases ($\chi = 0$, eg FHSV,
octonionic magic \cite{MAGIC})}

\item{$\cN =4=0_{_L} + 4_{_R}$, $\cN =2=0_{_L} + 2_{_R}$, $\cN
=1=0_{_L} + 1_{_R}$ NO massless R-R graviphotons, yet massless R-R
vectors couple to `exotic' D-branes }
\end{itemize}

Let us then sketch the $\cN =5,3$ cases.

\subsection{$\cN =5=1_{_L} + 4_{_R}$ case}

The simple(st) realization \cite{Ferrara:1989nm} is in terms of
$Z^L_2\times Z^L_2$ projections acting by T-duality along
$T^4_{6789}$ and $T^4_{4589}$ combined with order two shifts. One
can then identify 1/5 BPS bound states of D-branes carrying the
$8_{_{R-R}} =6_{(1533)} + 2_{(3333)}$ surviving R-R charges
(invariant orbits) $$ q^I_{(1335)}= q_1^{I} + {1\over
4!}\varepsilon_{i_Ij_Ik_Il_I} q_5^{Ii_Ij_Ik_Il_I} + {1\over
3!}\varepsilon^I{}_{J,K'L'} q_3^{JK'L'} + {1\over
3!}\varepsilon^I{}_{J,K"L"} q_3^{JK"L"} $$ where $i_I,j_I,k_I,l_I$
run over the four directions orthogonal to $T^2_I$ while $K',L'$
and $K", L"$ run over the two sets of two directions orthogonal to
$T^2_I$ and $$ q_{(3333)}^{I_1I_2I_3} = q_{3}^{I_1I_2I_3} +
{1\over 2!}\varepsilon^{I_2I_3}{}_{J_2J_3} q_3^{I_1J_2J_3} +
{1\over 2!}\varepsilon^{I_3I_1}{}_{J_3J_1} q_3^{J1I_2J_3} +
{1\over 2!}\varepsilon^{I_1I_2}{}_{J_1J_2} q_3^{J_1J_2I_3}$$

Other realizations of $\cN =5=1_{_L} + 4_{_R}$ are possible, all
lead invariably to the unique $\cN = 5$ supergravity massless
spectrum and LEEA: graviton $g_{\mu\nu}$ and 5 gravitini
$\psi_\mu$, 10 graviphotons $A_\mu$, 11 dilatini $\chi$ and 10
scalars $\phi$. The latter parameterize $\cM^{D=4}_{\cN = 5} =
SU(5,1)/U(5)$. $10_e + 10_m$ graviphotons in ${\bf 20}$ of
$SU(5,1)$ (3-index anti-symmetric tensor). The ``Minimal'' $\cN =
5$ superstring solutions have been classified into four classes
\cite{Ferrara:1989nm}. Two alternative superstring constructions
\cite{Dabholkar:1998kv} are available. A $Z_7$ asymmetric orbifold
of the $SU(7)$ torus $\theta_{_L} = (\omega_7,\omega_7^2,
\omega_7^4)$ and $7\sigma_{_R} =(1,2,-3,0,0,0,0)$ and a $Z_3$
asymmetric orbifold of the $SU(3)^3$ torus  $\theta_{_L} =
(\omega_3,\omega_3, \omega_3)$ and $3\sigma_{_R} =(1,-1,0;
1,-1,0;1,-1,0)$.

\subsection{$\cN =3=1_{_L} + 2_{_R}$ case}

The simplest $\cN=3$ model with 3 matter vector-plets can be
constructed in two steps. First one performs a `geometric' $Z_2$
freely acting orbifold (locally equivalent to $K3\times T^2$). The
$Z_2$ action combines a twist breaking $\cN=8=4_{_L} + 4_{_R}$ to
$\cN=4=2_{_L} + 2_{_R}$ and a shift preventing massless twisted
states.  Then a non geometric chiral (say Left-) projection
combined with a shift along the orthogonal directions $a=4,5$
breaks half of the $\cQ_L$. The surviving NS-NS charges are
$p_{_R}^a$ and their magnetic duals $\hat{P}_{_R}^a$. The
surviving R-R charges are the T-duality invariant combinations \be
q_1^a+{1\over 3!} \varepsilon^a_{bij} q_3^{bij} \quad , \quad
q_3^{aij} + {1\over 3!} \varepsilon^a_{bkl} q_5^{bijkl}\nn\ee At
most bound-states of the above can be 1/3 BPS states. No 1/2 BPS
states are allowed in $\cN=3$ supergravity. In particular the
massive gravitino in $\cN=4 \rightarrow \cN=3$ belong to a long
multiplet.

More $\cN =3=1_{_L} + 2_{_R}$ cases and otherwise can be found. A
complete classification of ``minimal'' $Z_2\times Z_2^L$ $\cN =3$
superstring solutions \cite{Ferrara:1989nm} lead to four classes
with $3+4K$ matter vector multiplets, with $K=0,1,2$, and eleven
sub-classes. Adding an extra chiral projection (ie splitting
geometric $Z_2$ into two chiral $Z_2$) produces models with $1+2K$
matter vector multiplets, with $K=0,1,2$. In particular a model
can be found with only one vector multiplet, thus having only
three complex scalar moduli (including dilaton!). Another
construction is based on the asymmetric $Z_3$ projection with
$\theta = (\omega_3,\omega_3, \omega_3; 1, \omega_3,
\omega_3^{-1})$ acting on the lattice of $SU(3)^3$.

Due to the rigidity of the LEEA and thus the scalar geometry $U(3,
N_v)/U(3)\times U(N_v)$ one expects duality with $\cN =3$
unoriented D3-brane models with 3-form fluxes \cite{Frey:2002hf}.

\subsection{Boundary states for L-R asymmetric branes}

In order to identify the open string excitations of the `exotic'
D-branes whose existence we have argued for above, one can resort
to the boundary state formalism \cite{Abouelsaood:1986gd,
Bianchi:2005sa, Billo:1998vr, Callan:1987px, Di Vecchia:1997pr, Di
Vecchia:2006gg}.

The boundary state can be factorized into a bosonic part and a
fermionic part. For the bosonic coordinates one has \be
|B_a\rangle^{(X)} = \sqrt{\det ({\cal G}_a + {\cal F}_a)} \exp (-
\sum_{n>o} a^i_{-n} R_{ij}(F_a) \tilde{a}^j_{-n}) |0_a\rangle
\nn\ee where $ R_a = (1 - F_a) / (1 +F_a) $ and $|0_a\rangle
\leftrightarrow p_{_L} = - R_a p_{_R}$ For the fermions in the
NS-NS sector (where no fermionic zero-modes are present) one finds
\be |B_a, \pm \rangle_{_{NS-NS}}^{(\psi)} = \exp (\pm i\sum_{n\ge
1/2} \psi^i_{-n} R_{ij}(F_a) \tilde{\psi}^j_{-n}) |\pm\rangle
\nn\ee while in the R-R sector one has \be |B_a, \pm
\rangle_{_{R-R}}^{(\psi)} = {1 \over \sqrt{\det ({\cal G}_a +
{\cal F}_a)}} \exp (i\pm \sum_{n>0} \psi^i_{-n} R_{ij}(F_a)
\tilde{\psi}^j_{-n}) {\cal U}^{\pm}_{A\tilde{B}}(F_a)
|A,\tilde{B}\rangle \nn\ee \be {\cal U}^{\pm}_{A\tilde{B}}(F_a) =
\left[{\rm AExp} (- F^a_{ij}\Gamma^{ij}/2) C\Gamma_{11} {1  \pm i
\Gamma_{11}\over 1 \pm i}\right]_{A\tilde{B}}\quad . \nn\ee

One can compute the partition functions (direct loop channel) and
determine the closed string couplings in the transverse tree
channel, where the amplitude is simply given by the overlap of the
boundary states. With these techniques magnetized and/or
intersecting D-branes in L-R symmetric orbifolds have been
described \cite{flux}. Our aim here is to generalize to
$Z^L_{N_{_L}}\times{Z^R_{N_{_R}}}$ action, and construct invariant
boundary states of the form $$ |B, F\rangle_g = {1\over
\sqrt{N_{_L} N_{_R}}} \left(1 + g_{_L} + g_{_R} + .... +
g^{N_{_L}-1}_{_L} g^{N_{_R}-1}_{_R}\right) |B, F\rangle = {1\over
\sqrt{N_{_L} N_{_R}}} \sum_{l,r} |B,F_{(l,r)}\rangle $$ where the
`induced' magnetic field $F_{(l,r)}$ is determined by the
condition $ R(F_{(l,r)}) = R(g^l_{_L}) R(F) R^t(g^r_{_R}) $.

The annulus amplitude then reads \be \cA_{g, h} = \Lambda(g,h)
\cI(g,h) \sum_\a c_\a^{^{GSO}}{\vartheta_\a(0)\over \eta^3}
\prod_I {\vartheta_\a(\epsilon_I(g,h)\tau)\over
\vartheta_1(\epsilon(g,h)\tau)} \nn\ee where $g=g_{_L} g_{_R}$ ,
$h=h_{_L} h_{_R}$, $\Lambda(g,h)$ is the lattice invariant under
$gh=g_{_L} h_{_L} g_{_R} h_{_R}$, while $\cI(g,h)$ is the
`intersection' number counting the invariant discrete zero-modes
and finally $\epsilon_I(g,h)$ are related to the eigenvalues of
$gh=g_{_L} h_{_L} g_{_R} h_{_R} \rightarrow
diag(e^{2i\epsilon_I(g,h)})$.

As an example consider the $\cN=5$ model based on the $T^6/Z^L_3$
torus of $SU(3)^3$. Prior to twists and shifts, there are 27
boundary states, associated to the `integrable' representations of
the current algebra at level $\kappa =1$, $ \cA_{\vec{r},\vec{s}}
= N_{\vec{r}, \vec{s}}^{\vec{t}} \cX_{\vec{t}} $, where
$\cX_{\vec{t}} = (V_8-S_8) \chi_{t_1}\chi_{t_2}\chi_{t_3}$
correspond to D-branes with magnetic quantum number $(n,m) =
(1,0), (-1,1), (0,-1)$. After twist and shift, invariant states
consist of branes rotated and displaced wrt one another \be
\cA_{Z_3^{L\neq R}} = {1\over 6} \sum_{a,b \in Z_3}
\Lambda_{(a,b)} \cI_{(a,b)} \sum_\a c_\a^{^{GSO}}
{\vartheta_\a(0)\over \eta^3} \prod_I {\vartheta_\a(a\tau +
b)\over \vartheta_1(a\tau + b)} \nn\ee As a result both
`untwisted' and `twisted' strings are present
\cite{Bianchi:1999uq, Blumenhagen:2000wh}, leading to a non chiral
spectrum in this case.

\section{Outlook}

Let us summarize what we said and speculate on possible
interesting developments of our analyses.

We hope we have convinced the audience, that twists and shifts
produce (unoriented) models with few or no T's that can
incorporate other mechanisms of moduli stabilization, \eg open
\cite{Antoniadis:2004pp, Dudas:2005jx, Bianchi:2005yz,
Bianchi:2005sa, Antoniadis:2005nu} and closed string fluxes
\cite{flux}, (non) anomalous $U(1)$'s \cite{Berkoozetal,
Bianchi:2007rb}, instanton effects, ... . Explicit computations
are feasible, though very systematic scans may be very
time-consuming (thousands of characters, ...)
\cite{Dijkstra:2004cc}. We have found a L-R symmetric `minimal'
model with only 1+1 twisted moduli \cite{Anastasopoulos:2009kj}
that escaped previous scans \cite{Donagi:2008xy, Kiritsis:2008mu}
but ... a perturbative L-R symmetric Type II model with
``$h_{11}$'' = ``$h_{21}$'' = 0 is yet to be found! Discrete
`deformations' (B-field etc) play a subtle role and can widen
phenomenological perspectives and, hopefully, reduce some
`tension' we observed between chirality and moduli stabilization

D-branes in L-R asymmetric vacua, e.g. $\cN=1_L + 0_R$, offer new
possibilities \cite{Bianchi:2008cj}. The relation with (non)
geometric fluxes is yet to be understood and interacting CFT's
(WZW, Gepner or alike) to be explored. In particular, abstract
SCFT with one-loop (super) characters $\cX_0$ (identity),
$\cX_{i}$ (massless chiral), $\cX^c_{i}$ (massless anti-chiral),
$\cX_I$ (massive $h_I>1/2$) may admit `exotic' modular invariants
of the form $$\cT_{IIB} = |\cX_0|^2 + \sum_i [\cX_{i}
\bar\cX_{I(i)} + \cX^c_{i} \bar\cX^c_{I(i)} + \bar\cX_{i}
\cX_{I(i)}  + \bar\cX^c_{i} \cX^c_{I(i)} ] + ... $$ where $I(i)$
labels massive characters with $h_{I(i)} = 1/2 + n_I$ (if
present). All moduli except the dilaton (multiplet) would be
`stabilized'. One cannot do any better for perturbative strings in
Minkowski space. So far, we have not been able to find a L-R
symmetric model of this kind but we will endure since we are
unaware of any no-go theorem and the `minimal' DJK model
\cite{Dolivet:2007sz}, though L-R asymmetric, offers good hopes.

Alternatively, one can start from Type I / Heterotic on $T^4/Z_2$
\cite{MBthesis, Berkoozetal}. No neutral twisted moduli are
present since they are eaten by anomalous $U(1)$'s), only
untwisted ones. Compactifying on $T^2$ and projecting by a freely
acting $Z_2$ can eliminate (some) untwisted moduli and produce no
extra twisted moduli. Including unoriented D-brane instantons
\cite{Bianchi:2007rb} and open string fluxes, one can hope to find
a `phenomenologically' viable model with all moduli stabilized by
mechanisms that are under full control of CFT techniques on the
worldsheet.

\section*{Acknowledgments}

 I would like to thank P.~Anastasopoulos, J.~F.~Morales Morera and G.~Pradisi
 for a very enjoyable collaboration on some of the topics covered in my talk.
 I would to thank the organizers of the 4-th RTN {\it
``Forces-Universe''} EU network Workshop in Varna,  and
M.~Chichmanova for the very kind hospitality. I delivered similar
talks at {\it Mathematical Challenges in String Phenomenology}
(ESI, Vienna, October 2008) and at {\it Fundamental Aspects of
Superstring Theory} (KITP, Santa Barbara, January 2009), where
this work has been completed. I would like to thank the organizers
for creating a stimulating environment. This work was supported in
part by MIUR-PRIN contract 2007-5ATT78, NATO PST.CLG.978785, RTN
grants MRTNCT- 2004-503369, EU MRTN-CT-2004-512194,
MRTN-CT-2004-005104 and in part by the National Science Foundation
under Grant No. PHY05-51164.



\begin{thebibliography}{99}

\bibitem{BPS}
  A.~Sagnotti,
  arXiv:hep-th/0208020.
  G.~Pradisi and A.~Sagnotti,
  Phys.\ Lett.\  B {\bf 216}, 59 (1989).
  M.~Bianchi and A.~Sagnotti,
  Phys.\ Lett.\  B {\bf 247} (1990) 517.
  M.~Bianchi and A.~Sagnotti,
  Nucl.\ Phys.\  B {\bf 361} (1991) 519.
For a review, see for instance C.~Angelantonj and A.~Sagnotti,
Phys.\ Rept.\  {\bf 371} (2002) 1 [Erratum-ibid.\  {\bf 376}
(2003) 339] [arXiv:hep-th/0204089].

\bibitem{Tfolds}
  M.~Dine and E.~Silverstein,
  arXiv:hep-th/9712166.

\bibitem{Narain:1986qm}
  K.~S.~Narain, M.~H.~Sarmadi and C.~Vafa,
  Nucl.\ Phys.\  B {\bf 288} (1987) 551.

\bibitem{Vafa:1986wx}
  C.~Vafa,
  Nucl.\ Phys.\  B {\bf 273} (1986) 592.
\bibitem{Kawai:1986va}
  H.~Kawai, D.~C.~Lewellen and S.~H.~H.~Tye,
  Phys.\ Rev.\ Lett.\  {\bf 57} (1986) 1832
  [Erratum-ibid.\  {\bf 58} (1987) 429],
  Phys.\ Rev.\  D {\bf 34} (1986) 3794,
  Nucl.\ Phys.\  B {\bf 288} (1987) 1.

\bibitem{abk}
I.~Antoniadis, C.~P.~Bachas and C.~Kounnas,
  Nucl.\ Phys.\  B {\bf 289} (1987) 87.
I.~Antoniadis and C.~Bachas,
  Nucl.\ Phys.\  B {\bf 298} (1988) 586.
\bibitem{Bianchi:2008cj}
  M.~Bianchi,
  Nucl.\ Phys.\  B {\bf 805} (2008) 168
  [arXiv:0805.3276 [hep-th]].
\bibitem{Lust:2008qc}
  D.~Lust, S.~Stieberger and T.~R.~Taylor,
  Nucl.\ Phys.\  B {\bf 808} (2009) 1
  [arXiv:0807.3333 [hep-th]].
\bibitem{Antoniadis:2007uz}
  I.~Antoniadis,
  arXiv:0710.4267 [hep-th];
\bibitem{Anastasopoulos:2009kj}
  P.~Anastasopoulos, M.~Bianchi, J.~F.~Morales and G.~Pradisi,
  arXiv:0901.0113 [hep-th].
\bibitem{Bianchi:1999uq}
  M.~Bianchi, J.~F.~Morales and G.~Pradisi,
  Nucl.\ Phys.\  B {\bf 573} (2000) 314
  [arXiv:hep-th/9910228].
\bibitem{Ferrara:1989nm}
S.~Ferrara and C.~Kounnas,
Nucl.\ Phys.\  B {\bf 328} (1989) 406.
\bibitem{ABKW}
  P.~Di Vecchia, V.~G.~Knizhnik, J.~L.~Petersen and P.~Rossi,
  Nucl.\ Phys.\  B {\bf 253}, 701 (1985).
I.~Antoniadis, C.~Bachas, C.~Kounnas and P.~Windey,
  Phys.\ Lett.\  B {\bf 171} (1986) 51.
\bibitem{CRISTINA}
  A.~E.~Faraggi, C.~Kounnas, S.~E.~M.~Nooij and J.~Rizos,
  Nucl.\ Phys.\  B {\bf 695}, 41 (2004)
  [arXiv:hep-th/0403058].
  A.~E.~Faraggi, S.~Forste and C.~Timirgaziu,
  JHEP {\bf 0608} (2006) 057
  [arXiv:hep-th/0605117].
  A.~E.~Faraggi, C.~Kounnas and J.~Rizos,
  Phys.\ Lett.\  B {\bf 648}, 84 (2007)
  [arXiv:hep-th/0606144].
  A.~E.~Faraggi, E.~Manno and C.~Timirgaziu,
  Eur.\ Phys.\ J.\  C {\bf 50} (2007) 701
  [arXiv:hep-th/0610118].
\bibitem{MBthesis}
M.~Bianchi, Ph. D Thesis, University of Rome ``Tor Vergata'',
1992; ROM2F-92-13
\bibitem{Dixon:1987yp}
L.~J.~Dixon, V.~Kaplunovsky and C.~Vafa,
Nucl.\ Phys.\  B {\bf 294} (1987) 43.
\bibitem{Donagi:2008xy}
R.~Donagi and K.~Wendland,
arXiv:0809.0330 [hep-th].
\bibitem{Camara:2007dy}
P.~G.~Camara, E.~Dudas, T.~Maillard and G.~Pradisi,
Nucl.\ Phys.\  B {\bf 795} (2008) 453 [arXiv:0710.3080 [hep-th]].
\bibitem{Dolivet:2007sz}
Y.~Dolivet, B.~Julia and C.~Kounnas,
JHEP {\bf 0802} (2008) 097 [arXiv:0712.2867 [hep-th]].
\bibitem{Vafa:1994rv}
  C.~Vafa and E.~Witten,
  J.\ Geom.\ Phys.\  {\bf 15} (1995) 189
  [arXiv:hep-th/9409188].
\bibitem{flux} For reviews, see e.g. M.~Grana,
  Phys.\ Rept.\  {\bf 423} (2006) 91
  [arXiv:hep-th/0509003];
  R.~Blumenhagen, B.~Kors, D.~Lust and S.~Stieberger,
  Phys.\ Rept.\  {\bf 445} (2007) 1
  [arXiv:hep-th/0610327],
and references therein.
\bibitem{MAGIC}
  M.~Bianchi and S.~Ferrara,
  JHEP {\bf 0802}, 054 (2008)
  [arXiv:0712.2976 [hep-th]].
\bibitem{BPStor}
  M.~Bianchi, G.~Pradisi and A.~Sagnotti,
  Nucl.\ Phys.\  B {\bf 376} (1992) 365.

 \bibitem{MBtor}
  M.~Bianchi,
  Nucl.\ Phys.\  B {\bf 528}, 73 (1998)
  [arXiv:hep-th/9711201].

\bibitem{EWtor}
  E.~Witten,
  JHEP {\bf 9802}, 006 (1998)
  [arXiv:hep-th/9712028].

\bibitem{Angelantonj:1999xf}
  C.~Angelantonj and R.~Blumenhagen,
  Phys.\ Lett.\  B {\bf 473} (2000) 86
  [arXiv:hep-th/9911190].

\bibitem{CBetal}
  C.~Bachas, M.~Bianchi, R.~Blumenhagen, D.~Lust and
T.~Weigand,
  JHEP {\bf 0808}, 016 (2008)
  [arXiv:0805.3696 [hep-th]].

\bibitem{Pesando:2008xt}
  I.~Pesando,
  Phys.\ Lett.\  B {\bf 668} (2008) 324
  [arXiv:0804.3931 [hep-th]].
\bibitem{Dabholkar:1998kv}
  A.~Dabholkar and J.~A.~Harvey,
  JHEP {\bf 9902}, 006 (1999)
  [arXiv:hep-th/9809122].

\bibitem{Kiritsis:2008mu}
  E.~Kiritsis, M.~Lennek and B.~Schellekens,
  arXiv:0811.0515 [hep-th].
\bibitem{Dai:1989ua}
  J.~Dai, R.~G.~Leigh and J.~Polchinski,
  Mod.\ Phys.\ Lett.\  A {\bf 4}, 2073 (1989).

\bibitem{Polchinski:1995mt}
  J.~Polchinski,
  Phys.\ Rev.\ Lett.\  {\bf 75}, 4724 (1995)
  [arXiv:hep-th/9510017].
\bibitem{Gimon:1996rq}
  E.~G.~Gimon and J.~Polchinski,
  Phys.\ Rev.\  D {\bf 54}, 1667 (1996)
  [arXiv:hep-th/9601038].
\bibitem{Berkooz:1996dw}
  M.~Berkooz and R.~G.~Leigh,
  Nucl.\ Phys.\  B {\bf 483} (1997) 187
  [arXiv:hep-th/9605049].
\bibitem{Abouelsaood:1986gd}
  A.~Abouelsaood, C.~G.~.~Callan, C.~R.~Nappi and S.~A.~Yost,
  Nucl.\ Phys.\  B {\bf 280}, 599 (1987).

\bibitem{Callan:1987px}
  C.~G.~.~Callan, C.~Lovelace, C.~R.~Nappi and S.~A.~Yost,
  Nucl.\ Phys.\  B {\bf 293}, 83 (1987).

\bibitem{Bianchi:2005sa}
  M.~Bianchi and E.~Trevigne,
  JHEP {\bf 0601}, 092 (2006)
  [arXiv:hep-th/0506080].

\bibitem{Di Vecchia:2006gg}
  P.~Di Vecchia, A.~Liccardo, R.~Marotta, F.~Pezzella and I.~Pesando,
  arXiv:hep-th/0601067.


\bibitem{Di Vecchia:1997pr}
  P.~Di Vecchia, M.~Frau, I.~Pesando, S.~Sciuto, A.~Lerda and R.~Russo,
  Nucl.\ Phys.\  B {\bf 507}, 259 (1997)
  [arXiv:hep-th/9707068].

\bibitem{Billo:1998vr}
  M.~Billo, P.~Di Vecchia, M.~Frau, A.~Lerda, I.~Pesando, R.~Russo and S.~Sciuto,
   ``Microscopic string analysis of the D0-D8 brane system and dual R-R
  Nucl.\ Phys.\  B {\bf 526}, 199 (1998)
  [arXiv:hep-th/9802088].
\bibitem{Brunner:1999fj}
  I.~Brunner, A.~Rajaraman and M.~Rozali,
  Nucl.\ Phys.\  B {\bf 558}, 205 (1999)
  [arXiv:hep-th/9905024].


\bibitem{Angelantonj:2000xf}
  C.~Angelantonj, R.~Blumenhagen and M.~R.~Gaberdiel,
   ``Asymmetric orientifolds, brane supersymmetry breaking and non-BPS
  Nucl.\ Phys.\  B {\bf 589}, 545 (2000)
  [arXiv:hep-th/0006033].

\bibitem{Gutperle:2000bf}
  M.~Gutperle,
  JHEP {\bf 0008}, 036 (2000)
  [arXiv:hep-th/0007126].

\bibitem{Gaberdiel:2002jr}
  M.~R.~Gaberdiel and S.~Schafer-Nameki,
  Nucl.\ Phys.\  B {\bf 654}, 177 (2003)
  [arXiv:hep-th/0210137].


\bibitem{Lawrence:2006ma}
  A.~Lawrence, M.~B.~Schulz and B.~Wecht,
  JHEP {\bf 0607}, 038 (2006)
  [arXiv:hep-th/0602025].

\bibitem{Kawai:2007qd}
  S.~Kawai and Y.~Sugawara,
  JHEP {\bf 0802}, 027 (2008)
  [arXiv:0709.0257 [hep-th]].
\bibitem{Frey:2002hf}
  A.~R.~Frey and J.~Polchinski,
  Phys.\ Rev.\  D {\bf 65}, 126009 (2002)
  [arXiv:hep-th/0201029].
\bibitem{Blumenhagen:2000wh}
  R.~Blumenhagen, L.~Goerlich, B.~Kors and D.~Lust,
   ``Noncommutative compactifications of type I strings on tori with  magnetic
  JHEP {\bf 0010}, 006 (2000)
  [arXiv:hep-th/0007024].
\bibitem{Antoniadis:2004pp}
  I.~Antoniadis and T.~Maillard,
  Nucl.\ Phys.\  B {\bf 716} (2005) 3
  [arXiv:hep-th/0412008].

\bibitem{Dudas:2005jx}
  E.~Dudas and C.~Timirgaziu,
  Nucl.\ Phys.\  B {\bf 716} (2005) 65
  [arXiv:hep-th/0502085].

\bibitem{Bianchi:2005yz}
  M.~Bianchi and E.~Trevigne,
  JHEP {\bf 0508} (2005) 034
  [arXiv:hep-th/0502147].

\bibitem{Antoniadis:2005nu}
  I.~Antoniadis, A.~Kumar and T.~Maillard,
  arXiv:hep-th/0505260.
  Nucl.\ Phys.\  B {\bf 767}, 139 (2007)
  [arXiv:hep-th/0610246].
\bibitem{Berkoozetal}
  M.~Berkooz, R.~G.~Leigh, J.~Polchinski, J.~H.~Schwarz, N.~Seiberg and E.~Witten,
  Nucl.\ Phys.\  B {\bf 475}, 115 (1996)
  [arXiv:hep-th/9605184].
\bibitem{Bianchi:2007rb}
  M.~Bianchi and J.~F.~Morales,
  JHEP {\bf 0802}, 073 (2008)
  [arXiv:0712.1895 [hep-th]].
\bibitem{Dijkstra:2004cc}
  T.~P.~T.~Dijkstra, L.~R.~Huiszoon and A.~N.~Schellekens,
  Nucl.\ Phys.\  B {\bf 710}, 3 (2005)
  [arXiv:hep-th/0411129].


\end{thebibliography}
\end{document}